\documentclass[aip,jmp,12pt]{revtex4-1}

\pdfoutput=1
\usepackage{amsmath}
\usepackage{amsfonts}
\usepackage[pdftex]{graphicx}
\usepackage{mathrsfs}
\usepackage{rotating}
\usepackage{setspace}
\usepackage{supertabular}
\usepackage[margin=1.4in]{geometry}
\usepackage{amsthm}

\begin{document}

\newtheorem{Theorem1}{Theorem}

{\Large \bf \noindent Spherical codes, maximal local packing density, and the golden ratio} \\
\noindent {\bf Adam B. Hopkins}$^{1}$, {\bf Frank H. Stillinger}$^{1}$ and {\bf Salvatore Torquato}$^{1-5}$ \\
$^1${\it \small Department of Chemistry, Princeton University, Princeton, New Jersey 08544} \\
$^2${\it \small Program in Applied and Computational Mathematics, Princeton University, Princeton, New Jersey 08544} \\
$^3${\it \small Princeton Center for Theoretical Science, Princeton University, Princeton, New Jersey 08544} \\
$^4${\it \small Princeton Institute for the Science and Technology of Materials, Princeton University, Princeton, New Jersey 08544} \\
$^5${\it \small School of Natural Sciences, Institute for Advanced Study, Princeton University, Princeton, New Jersey 08544}
\vfill
\noindent The densest local packing (DLP) problem in $d$-dimensional Euclidean space ${\mathbb R}^d$ involves the placement of $N$ nonoverlapping spheres of unit diameter near an additional fixed unit-diameter sphere such that the greatest distance from the center of the fixed sphere to the centers of any of the $N$ surrounding spheres is minimized. Solutions to the DLP problem are relevant to the realizability of pair correlation functions for sphere packings and might prove useful in improving upon the best known upper bounds on the maximum packing fraction of sphere packings in dimensions greater than three. The optimal spherical code problem in ${\mathbb R}^d$ involves the placement of the centers of $N$ nonoverlapping spheres of unit diameter onto the surface of a sphere of radius $R$ such that $R$ is minimized. It is proved that in any dimension, all solutions between unity and the golden ratio $\tau$ to the optimal spherical code problem for $N$ spheres are also solutions to the corresponding DLP problem. It follows that for any packing of nonoverlapping spheres of unit diameter, a spherical region of radius less than or equal to $\tau$ centered on an arbitrary sphere center cannot enclose a number of sphere centers greater than one more than the number that can be placed on the region's surface.

\newpage

\section{INTRODUCTION}

\begin{figure}[ht]
\centering
\includegraphics[width = 5.5in,viewport = 5 5 825
450,clip]{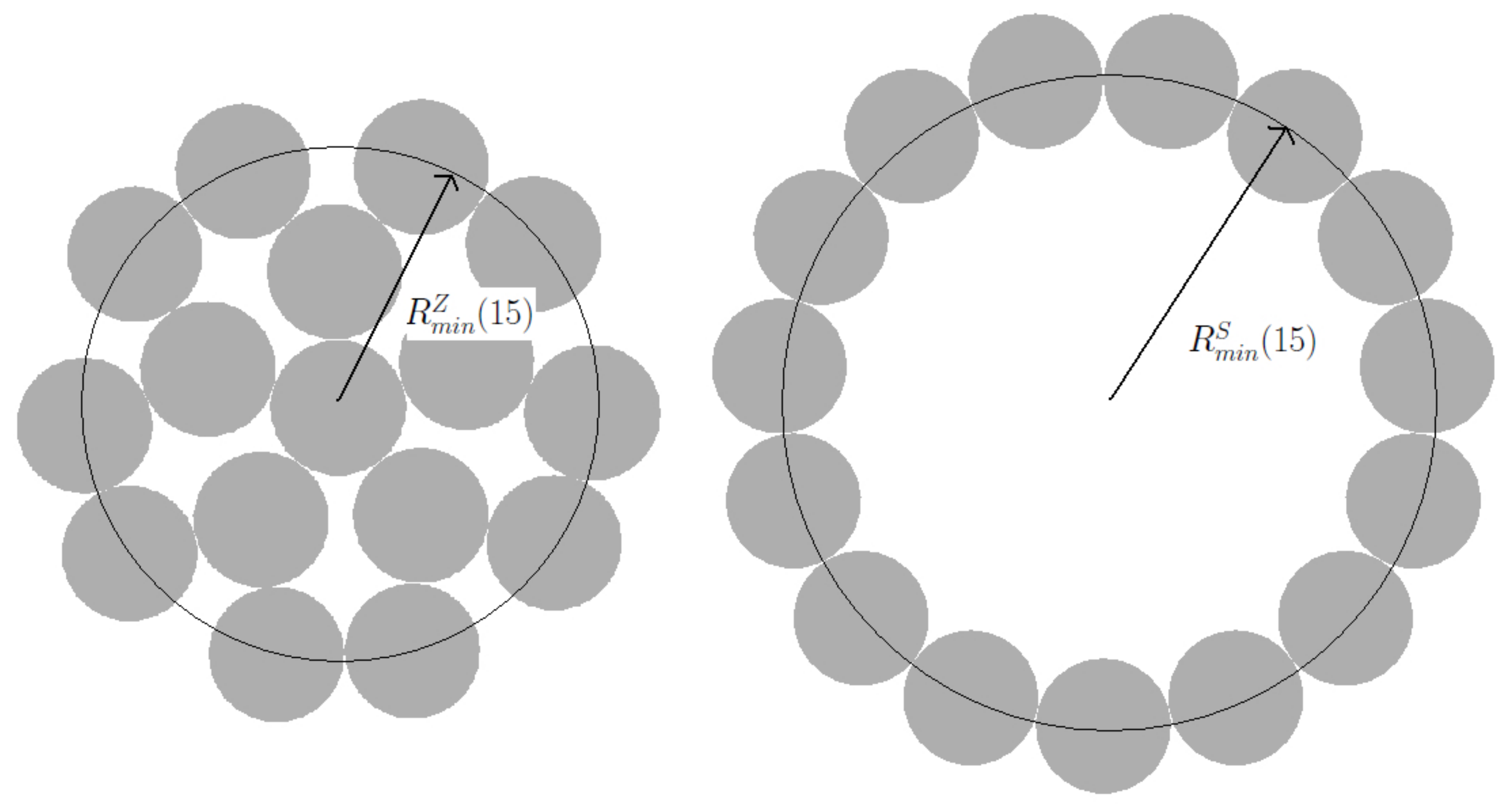}
\caption{Left: a conjectured optimal DLP configuration for $N=15$, $d=2$, $R^Z_{min}(15) = 1.873123\dots$. Right: a spherical code optimal configuration for $N=15$, $d=2$, $R^S_{min}(15) = 1/\big(2\sin{(\pi/15)}\big) = 2.404867\dots$.} 
\label{ZmaxVSsc15}
\end{figure}

The densest local packing (DLP) problem in ${\mathbb R}^d$ seeks an arrangement of $N$ spheres of unit diameter near (local to) an additional fixed central sphere such that the greatest radius $R$ between the centers of the surrounding $N$ spheres and the center of the central sphere is minimized. For an optimal configuration of $N$ spheres, i.e., a configuration for which $R$ is minimized, we call the minimized greatest radius $R^Z_{min}(N)$. The `$Z$' in the notation $R^Z_{min}(N)$ serves to distinguish from $R^S_{min}(N)$, where $R^S_{min}(N)$ in the optimal spherical code (OSC) problem is the radius of the minimal radius sphere onto the surface of which can be placed the centers of $N$ nonoverlapping spheres of unit diameter. For $N=15$, $d=2$, Fig. \ref{ZmaxVSsc15} depicts a conjectured optimal configuration for the DLP problem with minimal radius $R^Z_{min}(15) = 1.873123\dots$ alongside an optimal spherical code configuration with minimal distance $R^S_{min}(15) = 2.404867\dots$.

The kissing number problem in ${\mathbb R}^d$ seeks the maximum number $K_d$ of nonoverlapping spheres that may simultaneously be in contact with a (additional) sphere;\cite{CSSPLG1999} it is a special case of the DLP problem in that $K_d$ is equal to the greatest $N$ for which $R^Z_{min}(N) = 1$. The DLP problem can also be said to encompass the sphere packing problem in that in the limit as $N \rightarrow \infty$, optimal sphere packings and optimal DLP packings are equivalent. 

The maximum possible $N$ with $R = R^Z_{min}(N)$ for an optimal DLP configuration of $N$ spheres in ${\mathbb R}^d$ is the maximum of the function $Z({\bf r}_i,R)$. The function $Z({\bf r}_i,R)$ is defined for packings of nonoverlapping spheres of unit diameter as the number of sphere centers that are within distance $R$ from a sphere center at position ${\bf r}_i$, with $i$ an index over all centers and where the value of $Z({\bf r}_i,R)$ does not count the sphere center at ${\bf r}_i$. For a statistically homogeneous packing, the maximum at fixed $R$ of $Z({\bf r}_i,R)$ is an upper bound on the maximum of the function $Z(R)$, where $Z(R)$ is the expected number of sphere centers within distance $R$ from any given sphere center, or equivalently the average of $Z({\bf r}_i,R)$ over all $i$. For a packing that is also statistically isotropic, $Z(R)$ can be related to the pair correlation function $g_2(r)$, a function proportional to the probability density of finding a separation $r$ between any two points and normalized such that it takes the value of unity when no spatial correlations are present, by
\begin{equation}
Z(R) = \rho s_1(1)\int_0^Rx^{d-1}g_2(x)dx,
\label{ZR}
\end{equation}
where $\rho$ is the constant number density of points and $s_1(r)$ is the surface area of a sphere of radius $r$ in $d$ dimensions,
\begin{equation}
s_1(r) = \frac{2\pi^{d/2}r^{d-1}}{\Gamma(d/2)}.
\label{s1}
\end{equation}

The optimal spherical code (OSC) and DLP problems are similar. A spherical code is defined for parameters $(d,N,t)$ as a set of $N$ vectors from the origin to points on $S^{d-1} \subset {\mathbb R}^d$ such that the inner product between any two distinct vectors is less than or equal to $t$. The OSC problem is to minimize $t$ given $N$, or to maximize $N$ given $t$. There have been a number of investigations into the optimality and uniqueness of specific spherical codes (for example, see Ericson and Zinoviev\cite{EZCES2001} and Cohn and Kumar\cite{CK2007a}), and into providing bounds on $N$ given $t$ and $d$.\cite{CSSPLG1999}

A spherical code may be represented by a packing of $N$ nonoverlapping spheres of unit diameter with centers distributed on the surface of a sphere of radius $R$. In this representation, the OSC problem for a given $N$ requires finding the minimum $R$, $R^S_{min}(N)$, such that no two spheres overlap, i.e., such that the distance between the centers of any two spheres is greater than or equal to unity.

The OSC problem formulated in terms of nonoverlapping spheres and the DLP problem differ for all $N$ where $R^S_{min}(N) \geq 1$ only in that the former restricts the placement of sphere centers to a subset of the space allowed in the latter. From this observation, it is clear that when there exists a configuration of spheres that is a solution to the DLP problem with minimal radius $R^Z_{min}(N)$ that is also a spherical code, it is additionally a solution to the corresponding OSC problem, with $R^S_{min}(N) = R^Z_{min}(N)$. 

\section{THE DENSEST LOCAL PACKING PROBLEM AND REALIZABILITY}

Only functions obeying certain necessary conditions known as realizability conditions can be correlation functions of point processes in ${\mathbb R}^d$.\cite{Lenard1975a,TS2002a,KLS2007a} Two realizability conditions on the pair correlation function $g_2(r)$ are the nonnegativity of $g_2(r)$ and its corresponding structure factor $S(k)$ at all points $r$ and $k$.\cite{TS2002a} These two conditions appear to be strong conditions for the realizability of sphere packings (point processes in which the minimum pair separation distance is unity), especially as the space dimension increases.\cite{TS2006a} They have been employed, among other uses, to provide conjectures for a lower bound on the maximum packing fraction of an infinite sphere packing in any dimension,\cite{TS2006a} and to demonstrate the feasibility in three dimensions of a sequence of disordered packings whose disorder vanishes as density approaches the maximum possible.\cite{HST2009a}

Cohn and Elkies\cite{CE2003a} employ analogs of these two conditions, in conjunction with a linear programming technique, to find the best known bounds on the packing fraction of infinite sphere packings in (at least) dimensions four through 36. In the conclusions of a previous work,\cite{HST2009a} we discuss how a third realizability condition, found by solving the DLP problem for a packing of 13 spheres in three dimensions, can improve upon the three-dimensional bound found in Ref. 9.

The technique employed in Ref. 7 to find conjectured lower bounds has been shown to be the dual of the primal infinite-dimensional linear program employed in Ref. 9, and Cohn and Kumar\cite{CK2009a} have since shown that there is no duality gap between the two programs. This means that when the best $g_2(r)$ test functions are employed, the upper and conjectured lower bounds will coincide. Cohn and Elkies in Ref. 9 were able to find a test function that yields the best upper bound on the maximal packing fraction in three dimensions, a packing fraction of $0.778$, which is well above the true maximum. This means that there is a test function for the lower bound formulation that will deliver the same packing fraction of $0.778$, which is clearly not realizable.

A putative improvement on the upper bound in ${\mathbb R}^3$ was obtained by employing an estimate for $R^Z_{min}(13)$ in the DLP problem in ${\mathbb R}^3$.\cite{CKT2009a} Requiring that $Z(R) \leq 12$ up to some small positive $\alpha$ beyond contact, with $R = 1 + \alpha$ the estimate for $R^Z_{min}(13)$, reduces the $d=3$ bound in Ref. 9. For example, estimating $\alpha = 0.05$\cite{footnote1} reduces the bound from $0.778$ to $0.771$. This result strongly suggests that DLP solutions introduce more information than is contained in the pair correlation function alone, in that there is at least one test $g_2(r)$ that obeys the two nonnegativity conditions but violates the bound $Z(1 + \alpha) \leq 12$.

Further solutions to the DLP problem provide additional realizability conditions that might be employed to improve upon the upper bounds on infinite sphere packings in dimensions greater than three. For a statistically homogeneous and isotropic packing of spheres, these additional conditions may be written as
\begin{equation}
Z(R) \leq Z_{max}(R),
\label{ZRbound}
\end{equation}
where the function $Z_{max}(R)$ is defined in ${\mathbb R}^d$ as the maximum number of sphere centers that fit within distance $R$ from a central sphere center.\cite{footnote2} It is clear that $Z_{max}(R)$ in ${\mathbb R}^d$ is completely defined by the solutions to the densest local packing problem at all $N$. 

In the following section, we show that any configuration of $N$ $d$-dimensional spheres near a (additional) sphere fixed at the origin, with the greatest of the $N$ distances from the origin to the $N$ sphere centers equal to $R \leq \tau = (1+\sqrt{5})/2 \approx 1.618034$ the golden ratio, may be transformed to a spherical code in the sense of nonoverlapping spheres, also of radius $R$. As this statement is applicable to any configuration of $N$ spheres that is a solution to the DLP problem with $R \leq \tau$, it follows that any optimal spherical code with radius $R^S_{min}(N) \leq \tau$ is also an optimal configuration for the corresponding DLP problem, with $R^Z_{min}(N) = R^S_{min}(N)$.

\section{TRANSLATING UNIT-DIAMETER SPHERES TO THE SURFACE AT RADIUS $R \leq \tau$}

The key idea behind the proof of the above statement involves translating sphere centers radially outward to a spherical surface of radius $R$. The idea of radially translating points to a spherical surface has been employed by Melissen\cite{Melissen1994a} to aid a proof of the optimality of certain packings of 11 congruent nonoverlapping circles in a circle and more recently by Cohn and Kumar\cite{CK2009b} to rescale vectors in ${\mathbb R}^{24}$ to terminate on $S^{23}$. However, prior to this work, the maximum radius from the center of a fixed nonoverlapping sphere to which the centers of surrounding spheres can be translated without resulting overlap was not known.

Specifically, for any number of nonoverlapping spheres of unit diameter initially situated such that their centers are contained in a spherical shell of radial span $[1,R]$ with $1 \leq R \leq \tau$, all sphere centers at a distance less than $R$ from the center of the shell may be translated radially outward to distance $R$ without any resulting overlap between spheres. This statement more generally applies (via a simple scaling argument) to congruent nonoverlapping spheres of arbitrary diameter $D$ that are contained within a spherical shell of radial span $[D,R]$, $D \leq R \leq \tau D$.

Define ${\mathbb A}_N(R)$ in ${\mathbb R}^d$ as the set of all packings of any number $N$ of nonoverlapping spheres of unit diameter with centers situated in a spherical shell of radial span $[1,R]$, with $R$ the greatest of the distances from the center of the shell (the origin) to the $N$ sphere centers. An element of the set ${\mathbb A}_N(R)$ represents any arrangement of $N$ spheres with greatest distance $R$ situated near an additional nonoverlapping sphere fixed at the origin.

\begin{Theorem1}
Consider any single element of ${\mathbb A}_N(R)$ in ${\mathbb R}^d$. For $R \leq \tau$, all $N$ spheres may be translated radially outward such that their centers are at distance $R$ from the origin and still remain an element of ${\mathbb A}_N(R)$. For $R > \tau$, $d > 1$, there exist elements of ${\mathbb A}_N(R)$ such that an outward radial translation of a given sphere center to distance $R$ will yield overlap between at least two of the $N$ spheres.
\end{Theorem1}

\begin{proof}
The proof proceeds from the law of cosines in the method of the proof of Lemma 4.1 in Ref. 15. For any two of the $N$ spheres with centers situated at distances $b$, $c$ from the origin and separated by distance $a$, the cosine of the angle formed between the two centers at the origin, taken such that $0 \leq \theta \leq \pi$, is
\begin{equation}
\cos{\theta} = \frac{b^2 + c^2 - a^2}{2bc}.
\label{cosTheta}
\end{equation}
For nonoverlapping spheres of unit diameter, $a \geq 1$, and
\begin{equation}
\cos{\theta} \leq \frac{b^2 + c^2 - 1}{2bc},
\label{cosThetaContact}
\end{equation}
where the equality holds when the two spheres are in contact.

Over the range $b \geq 1$, $c \geq 1$, the function $\cos{\theta}$ in Eq. (\ref{cosThetaContact}) is convex individually in both $b$ and $c$. This implies that $\cos{\theta}$ must be at a maximum at one of the corners of the square $1 \leq b \leq R$, $1 \leq c \leq R$. If $R > \tau$, the point $(1,R)$ (or equivalently $(R,1)$) yields the maximum, whereas for $R < \tau$, the point $(R,R)$ yields the maximum, with $(1,R)$ and $(R,R)$ both yielding the maximum at $R=\tau$.

It follows directly that for $R \leq \tau$, the minimum possible angle at the origin between any two of the centers of $N$ spheres that are an element of ${\mathbb A}_N(R)$ is the angle present when two of the centers are placed at distance $R$ from the origin and distance unity from one another. An outward radial translation of one or both of any pair of centers to distance $R$ will therefore yield no overlap between the two spheres, as the angle between the centers must be greater than or equal to the angle present when two spheres are in contact with one another with centers at distance $R$ from the origin. As this holds for any pair of the $N$ sphere centers, all centers at a distance less than $R$, $R \leq \tau$, may be translated radially outward to distance $R$ without any resulting overlap.

For $R > \tau$, $d > 1$, overlap between two spheres is possible after an outward radial translation. For example, when two spheres are initially in contact with centers at distance unity from each other and at distances $1$ and $R > \tau$ from the origin, the angle formed at the origin between centers is smaller than the angle present when the spheres are in contact with centers both at distance $R$. A radial translation outward of the sphere center at distance $1$ to distance $R$ would thus yield overlap. This concludes the proof of Theorem 1.
\end{proof}

\section{RESULTS AND DISCUSSION}

Theorem 1 applies to any configuration of $N$ nonoverlapping spheres that are an element of ${\mathbb A}_N(R)$. In particular, in an optimal DLP configuration for $N$ spheres in ${\mathbb R}^d$ with $R^Z_{min}(N) \leq \tau$, which is by definition an element of ${\mathbb A}_N(R^Z_{min}(N))$, any of the spheres with centers not at distance $R^Z_{min}(N)$ from the origin may be translated radially outward to distance $R^Z_{min}(N)$ without any overlap between spheres. The resulting configuration is both a solution to the DLP problem and, in the sense of nonoverlapping spheres, to the corresponding OSC problem, with $R^S_{min}(N) = R^Z_{min}(N)$. Theorem 1 therefore implies that while for $1 \leq R^Z_{min}(N) \leq \tau$ there may be solutions to the densest local packing problem that are not spherical codes, for $1 \leq R^S_{min}(N) \leq \tau$ there are no solutions to the optimal spherical code problem that are not also solutions to the corresponding densest local packing problem. 

The kissing numbers $K_d$ in ${\mathbb R}^d$ are only known rigorously for $d=1\,.\,.\,4$, $d=8$ and $d=24$;\cite{Musin2008a,CSSPLG1999} for $d =1$, $2$, $3$ and $4$, they are $2$, $6$, $12$ and $24$,\cite{Musin2008a} respectively. For $N \leq K_d$, the solution to the DLP problem is simply $R^Z_{min}(N) = 1$ by necessity as the nonoverlapping sphere of unit diameter at the origin is fixed. For $N$ such that $K_d < N \leq N^{\tau}_d$, where we define $N^{\tau}_d$ in ${\mathbb R}^d$ as the greatest integer $N$ such that $R^S_{min}(N) \leq \tau$, the optimal spherical codes are solutions to the corresponding DLP problems with $R^Z_{min}(N) = R^S_{min}(N)$. The questions concerning the values of $N^{\tau}_d$ in each dimension and how $N^{\tau}_d$ grows with $d$ naturally emerge.

In one dimension, the answer to the first question is trivial, with $N^{\tau}_1 = 2$. In two dimensions, optimal spherical codes can be found analytically via simple trigonometry, with $R^S_{min}(N) = \tau$ for $N=10$, or $N^{\tau}_2 = 10$. Strong conjectured solutions for $R^S_{min}(N)$ that serve (at least) as upper bounds to the OSC problem are well-known in low dimensions greater than two for small $N$.\cite{SloaneSCweb} For $d=3$, these yield the conjecture $N^{\tau}_3 = 33$ with $R^S_{min}(33) \approx 1.607051$. For $d=4$, a unique optimal spherical code is known such that $R^S_{min}(120) = \tau$, giving the result that $N^{\tau}_4 = 120$.\cite{footnote3}

The question of precisely how $N^{\tau}_d$ grows with $d$ is still open, and is more complicated; however, bounds may be established via known bounds on $N$ (given $d$ and $t$) for optimal spherical codes, such as with those given in chapter two of Ref. 1. The lower bound (due to Wyner\cite{Wyner1965a}) on $N(d,\phi)$ for a spherical code of minimum angle $\phi = \cos^{-1}(t)$ in dimension $d$ is
\begin{equation}
N(d,\phi) \geq \frac{1}{\sin^d(\phi)},
\label{lowerSCbound}
\end{equation}
giving for $R^S_{min}(N) = \tau$, $t = \tau/2$ and $N_d^{\tau} \geq 1.7013^d$. This may be compared to the lower bound on the kissing number obtained from (\ref{lowerSCbound}), $K_d \geq 1.1547^d$.  The upper bound due to Rankin\cite{Rankin1955a} is
\begin{equation}
N(d,\phi) \leq \left(\frac{1}{2}\pi d^3\cos(\phi)\right)^{1/2}\left(\sqrt{2}\sin{(\phi/2)}\right)^{-d},
\label{upperSCbound}
\end{equation}
giving, for $R^S_{min}(N) = \tau$, $N_d^{\tau} \leq 1.1273d^{3/2} \times 2.2883^d$. This may be compared to the Kabatiansky-Levenshtein\cite{KL1978a} upper bound on the kissing number, $K_d \leq 1.3205^d$. Comparing the upper bound on the kissing number and the lower bound on $N_d^{\tau}$, it is clear that $N_d^{\tau}$ grows exponentially faster than $K_d$. 

\bigskip
\noindent {\bf ACKNOWLEDGEMENTS:}
\medskip

The authors thank Henry Cohn for valuable comments and suggestions concerning the manuscript. S.T. thanks the Institute for Advanced Study for its hospitality during his stay there. This work was supported by the Division of Mathematical Sciences at the National Science Foundation under Award Number DMS-0804431 and by the MRSEC Program of the National Science Foundation under Award Number DMR-0820341.




\newpage

\begin{thebibliography}{10}

\bibitem{CSSPLG1999}
Conway, J.~H. and Sloane, N. J.~A.
\newblock {\em Sphere Packings, Lattices and Groups}.
\newblock Springer,  (1999).

\bibitem{EZCES2001}
Ericson, T. and Zinoviev, V.
\newblock {\em Codes on Euclidean Spheres}.
\newblock North-Holland,  (2001).

\bibitem{CK2007a}
Cohn, H. and Kumar, A.
\newblock {\em New York J. Math.}{ \bf 13}, 147--157 (2007).

\bibitem{Lenard1975a}
Lenard, A.
\newblock {\em Arch. Ration. Mech. Anal.}{ \bf 59}, 219--239, 241--256 (1975).

\bibitem{TS2002a}
Torquato, S. and Stillinger, F.
\newblock {\em J. Phys. Chem. B}{ \bf 106}, 8354--8359 (2002).

\bibitem{KLS2007a}
Kuna, T., Lebowitz, J.~L., and Speer, E.~R.
\newblock {\em J. Stat. Phys.}{ \bf 129}, 417--439 (2007).

\bibitem{TS2006a}
Torquato, S. and Stillinger, F.~H.
\newblock {\em Exp. Math.}{ \bf 15}, 307--331 (2006).

\bibitem{HST2009a}
Hopkins, A.~B., Stillinger, F.~H., and Torquato, S.
\newblock {\em Phys. Rev. E}{ \bf 79}, 031123 (2009).

\bibitem{CE2003a}
Cohn, H. and Elkies, N.
\newblock {\em Ann. Math.}{ \bf 157}, 689--714 (2003).

\bibitem{CK2009a}
Cohn, H. and Kumar, A.
\newblock Unpublished.

\bibitem{CKT2009a}
Cohn, H., Kumar, A., and Torquato, S.
\newblock Unpublished.

\bibitem{footnote1}
The actual number $1 + \alpha$ is strongly conjectured to be $1.045573\dots$,
  equal to the current best lower bound for $R^S_{min}(13)$ in ${\mathbb R}^3$.

\bibitem{footnote2}
In the sense of $Z(R)$ defined in Eq. (\ref{ZR}) for a statistically
  homogeneous packing, $Z_{max}(R)$ is generally not a sharp upper bound for
  $Z(R)$, i.e., there is not always a configuration of spheres for which
  equality in (\ref{ZRbound}) holds. This is because $Z_{max}(R)$ is defined
  locally, in terms of one central sphere, whereas $Z(R)$ in Eq. (\ref{ZR}) is
  defined globally in terms of a probability density, or in the case of a
  finite packing, in terms of an average over all spheres.

\bibitem{Melissen1994a}
Melissen, H.
\newblock {\em Geom. Dedicata}{ \bf 50}, 15--25 (1994).

\bibitem{CK2009b}
Cohn, H. and Kumar, A.
\newblock {\em To appear in Ann. Math.}
\newblock \url{http://arxiv.org/abs/math.MG/0403263}

\bibitem{Musin2008a}
Musin, O.~R.
\newblock {\em Ann. Math.}{ \bf 168}, 1--32 (2008).

\bibitem{SloaneSCweb}
Sloane, N. J.~A., Hardin, R.~H., and Smith, W.~D.
\newblock \url{www.research.att.com/~njas/packings/}

\bibitem{footnote3}
It has been shown that the vertices of the $600$-cell are the unique
  $(4,120,\tau/2$ spherical code,\cite{Boroczky1978a,BD2001a} which corresponds
  in ${\mathbb R}^4$ to $R^S_{min}(120) = \tau$.

\bibitem{Wyner1965a}
Wyner, A.~D.
\newblock {\em Bell Sys. Tech. J.}{ \bf 44}, 1061--1122 (1965).

\bibitem{Rankin1955a}
Rankin, R.~A.
\newblock {\em Proc. Glasgow Math. Assoc.}{ \bf 2}, 139--144 (1955).

\bibitem{KL1978a}
Kabatiansky, G.~A. and Levenshtein, V.~I.
\newblock {\em Probs. of Info. Trans.}{ \bf 14}, 1--17 (1978).

\bibitem{Boroczky1978a}
B\"{o}r\"{o}czky, K.
\newblock {\em Acta Math. Acad. Sci. Hung.}{ \bf 32}, 243--261 (1978).

\bibitem{BD2001a}
Boyvalenkov, P. and Danev, D.
\newblock {\em Arch. Math.}{ \bf 77}, 360--368 (2001).

\end{thebibliography}

\end{document}